\documentclass[10pt]{article}
\usepackage{psfig,epsfig,amssymb} 
\usepackage{multicol}  
\input psfig.sty
\input cyracc.def
\input amssymb.sty
\usepackage{color}
\usepackage{bm}

\newfont{\yihao}{cmb10 at 18pt}

\DeclareSymbolFont{lettersA}{U}{txmia}{m}{it}
\DeclareMathSymbol{\piup}{\mathord}{lettersA}{25}
\DeclareMathSymbol{\muup}{\mathord}{lettersA}{22}      %
\renewcommand{\baselinestretch}{1.06} 
 \catcode`@=11

\let\@oddfoot\@empty  \let\@evenfoot\@empty
\def\@evenhead{\small\thepage\hfill {\small\it Fulvio MELIA, Front. Phys. }\hfill }
\def\@oddhead{\hfill{\small\it Fulvio MELIA, Front. Phys. }\hfill\small\thepage}

\definecolor{oranged}{cmyk}{0.0,0,0.0,0.5}
\newfont{\xbt}{cmb10 at 12pt}

\usepackage{graphicx}                 

\graphicspath{{figs/}}      
\newcommand\simlt{\lower.5ex\hbox{$\; \buildrel < \over \sim \;$}}
\newcommand\simgt{\lower.5ex\hbox{$\; \buildrel > \over \sim \;$}}

\begin{document}
\thispagestyle{empty}


\begin{center}

{\usefont{T1}{fradmcn}{m}{n}\yihao Physical Basis for the Symmetries in the Friedmann--Robertson--Walker Metric}
\vspace*{5mm}

{\bf\small Fulvio MELIA$^\dag$}\vspace*{4mm}\footnote{$^\dag$John Woodruff Simpson Fellow.}

{\footnotesize \it 
Department of Physics, The Applied Math Program, and Department of Astronomy,\\
The University of Arizona, Tucson, AZ 85721, USA\\[1.5mm]
E-mail: fmelia@email.arizona.edu\\[.5mm]
}
\vspace{5mm}

\baselineskip 10pt
\renewcommand{\baselinestretch}{0.8}
\parbox[c]{152mm}
{\noindent{Modern cosmological theory is based on the Friedmann--Robertson--Walker (FRW) metric. Often
written in terms of co-moving coordinates, this well-known solution to Einstein's equations
owes its elegant and highly practical formulation to the cosmological principle and Weyl's
postulate, upon which it is founded. However, there is physics behind such symmetries, and not all
of it has yet been recognized. In this paper, we derive the FRW metric coefficients from the
general form of the spherically symmetric line element and demonstrate that, because the
co-moving frame also happens to be in free fall, the symmetries in FRW are valid only for a medium
with zero active mass. In other words, the spacetime of a perfect fluid in cosmology may be
correctly written as FRW only when its equation of state is $\rho+3p=0$, in terms of the
{\it total} pressure $p$ and {\it total} energy density $\rho$. There is now compelling observational
support for this conclusion, including the Alcock--Paczy\'nski test, which shows that only an FRW
cosmology with zero active mass is consistent with the latest model-independent baryon
acoustic oscillation data.
\vspace{2mm}

 \vspace{2mm}

}}

\end{center}

\normalsize

\baselineskip 12pt \renewcommand{\baselinestretch}{1.06}
\parindent=10.8pt  \parskip=0mm \rm\vspace{5mm}

\begin{multicols}{2}
\setlength{\parindent}{1em}    

\vspace*{6.0mm} \hrule\vspace{2mm} 
\noindent {\large \usefont{T1}{fradmcn}{m}{n}\xbt 1\quad Introduction}\vspace{2.5mm}

\noindent The gravitational collapse (or expansion) of a spherically symmetric distribution of matter
and energy was first considered in Ref. [1]. Since then, several important generalizations 
have been made in Refs. [2--4], among others, each of which introduced essential physical 
ingredients, such as the influence of non-zero pressure.

Perhaps because the work of Birkhoff [5] had not yet been fully appreciated,
the development of what we now call the Friedmann--Robertson--Walker (FRW) metric took a
different approach from that of the general problem of gravitational expansion and
contraction [6]. The corollary to Birkhoff's theorem, however, states that even for an 
infinite, isotropic medium---be it dynamic or static---the spacetime within a spherical 
shell is independent of what lies beyond the enclosed spherical volume.$^\ddag$\footnote{$^\ddag$Weinberg
[7] provides an excellent, though brief, description of this phenomenon. See also
Ref. [8] for a more recent discussion.} It is therefore not difficult to convince oneself
that the general relativistic description of the universal expansion of a shell at radius $R$ relative
to an observer at the origin of the coordinates should closely mirror the formalism employed in
problems of stellar collapse or explosion involving a body of the same size.

The difference between the two approaches is highlighted by the fact that, whereas
the dynamical equations of gravitational collapse are obtained by solving Einstein's equations
with a general form of the metric, the Friedmann equations are derived after all the
symmetries have been used to greatly simplify the FRW metric {\it before} it is
introduced into the field equations. For example, the lapse function in this metric
is conventionally set equal to one without considering possible dilation effects
due to an accelerated expansion of the spatial coordinates. However, as we shall
show in this paper, such an approach bypasses at least one important condition
that the stress-energy tensor must satisfy in order to permit this simple choice
of metric. In so doing, we will demonstrate that the FRW spacetime is actually
valid only for a perfect fluid with zero active mass.

This critical re-evaluation of the applicability of the FRW spacetime to arbitrary
constituents of the cosmic fluid is motivated in large part by the ever-improving
precision of cosmological measurements [9], which are refining
our view of the cosmic equation of state. The current standard model is an FRW
cosmology with a relatively unconstrained blend of constituents, including
matter ($\rho_{\rm m}$), radiation ($\rho_{\rm r}$), and an unknown dark energy
($\rho_{\rm de}$), and their associated pressures, $p_{\rm m}$, $p_{\rm r}$,
and $p_{\rm de}$, respectively. The existence of $\rho_{\rm de}$ has been demonstrated
convincingly by meticulous analysis of Type Ia SN data, which prove beyond
any doubt that the expansion of the Universe is not slowing down, an otherwise
unavoidable outcome if matter and radiation were acting alone [10--13]. 
When dark energy is assumed to be a cosmological
constant, $\Lambda$, with $p_{\rm de}=-\rho_{\rm de}$, the model is referred
to as the $\Lambda$ cold dark matter ($\Lambda$CDM) model; otherwise, the
conventional designation is $w$CDM, where $w\equiv p_{\rm de}/\rho_{\rm de}$
characterizes the dark energy equation of state.

Over the past several decades, $\Lambda$CDM/$w$CDM has been quite successful
in accounting for the observations, thanks chiefly to the flexibility it enjoys owing to
a rather large number of free parameters. These include $w$; the Hubble constant $H_0$,
which represents the expansion rate today; the scaled matter ($\Omega_{\rm m}$)
and dark energy ($\Omega_{\rm de}$) densities, where $\Omega_i\equiv 
\rho_i/\rho_{\rm c}$ and $\rho_{\rm c}\equiv 3c^2{H_0}^2/8\pi G$ is the
so-called critical density; and the partitioning of $\rho_{\rm m}$ into baryonic
and dark matter. The scaled radiation energy density, $\Omega_{\rm r}\equiv
\rho_{\rm r}/\rho_{\rm c}$, is not considered to be a free parameter because
we can measure the temperature of the cosmic microwave background (CMB) 
(blackbody) radiation very accurately. All told, the standard model has at least
five unspecified parameters, all of which can be adjusted to fit the data.

Given the wide latitude permitted by this parametrization, optimization of
the model parameters by fitting the observations (most impressively through measurements
of the CMB [14--16]) is revealing a very surprising
result: over a Hubble time (i.e., ${H_0}^{-1}$), the Universe expanded by an
amount equal to the amount under constant expansion, despite 
the fact that the combination of $\rho_{\rm m}$, $\rho_{\rm r}$, and $\rho_{\rm de}$
should have produced periods of deceleration and acceleration. In other words,
the average acceleration of the Universe up to this point in time is zero within
the measurement errors. A more meaningful way to say this is that averaged
over a Hubble time, the quantity $p/\rho$, where $p=p_{\rm r}+p_{\rm m}
+p_{\rm de}$ and $\rho=\rho_{\rm r}+\rho_{\rm m}+\rho_{\rm de}$, yields
 $\langle p/\rho\rangle=-1/3$.

What makes this result even more striking is that, in the context  of $\Lambda$CDM,
the Universe is open and infinite. However, the condition $\langle p/\rho \rangle
=-1/3$ can be achieved only once in its entire (presumably infinite) history, and
it is happening right now, just when we are looking. Such an astonishing
coincidence begs for a physical explanation. By demonstrating that the symmetries
in FRW require zero active mass in the cosmic fluid, we will show in this paper
that $\rho+3p$ is in fact always zero, and that the result $\langle p/\rho\rangle=
-1/3$ is therefore independent of time. Instead, it must be an imperfect (or
incomplete)  parametrization of the standard model that leads to an inferred variable
expansion rate. Further, to maintain consistency with the condition $\langle p/\rho\rangle=-1/3$,
it is therefore the optimized parameter values that must change depending on when
the fits to the data are made.
 
\vspace*{6.0mm} \hrule\vspace{2mm} \noindent {\large
\usefont{T1}{fradmcn}{m}{n}\xbt 2\quad General Relativistic Expansion and Contraction}\vspace{2.5mm}

\noindent The FRW metric for a spatially homogeneous and isotropic three-dimensional space
is usually written as
\begin{eqnarray}
ds^2&=&c^2dt^2-a^2(t)\left[dr^2(1-kr^2)^{-1}+\right.\nonumber\\
&\null&\qquad\qquad\qquad\quad \left.r^2(d\theta^2+\sin^2\theta\,d\phi^2)\right]\;
\end{eqnarray}
in terms of the cosmic time $t$, co-moving radius $r$,
universal expansion factor $a(t)$, and angular coordinates $\theta$ and $\phi$ in
the co-moving frame. The spatial curvature constant $k$ takes the values $(-1,0,+1)$
for an open, flat, or closed universe, respectively.

Clearly, not only is the fluid at rest in this coordinate system $(ct,r,\theta,\phi)$, but
the corresponding frame must also be in free fall, because $g_{tt}=1$. To see this,
one need only remember Einstein's demonstration that his theory of gravity, based on
the equivalence principle, correctly reduces to Newton's law in the weak-field limit if
\begin{equation}
g_{tt}=1+{2\Phi\over c^2}\;,
\end{equation}
where $\Phi$ is the gravitational potential. Obviously, if $g_{tt}=1$, an observer in
this co-moving frame sees no gravity. Thus, we should immediately ask ourselves under what
conditions the ansatz in Eq.~(1) is justified when we apply it to cosmology, where
one typically assumes a perfect fluid with the stress-energy tensor
\begin{equation}
T_{\alpha\beta}=\left(\rho_m+{p\over c^2}\right)u_\alpha u_\beta-pg_{\alpha\beta}\;
\end{equation}
in terms of the co-moving energy density $\rho=\rho_m c^2$ (where $\rho_m$ is the
equivalent mass density), pressure $p$, and four-velocity
$u_\alpha$. For example, we might wonder whether the symmetries built into Eq.~(1)
place any constraints on the pressure, which in fact they do, as $p$
must be homogeneous and isotropic.

Let us now take a step backward and, instead of using the FRW metric as given
in Eq.~(1), treat it as a special case of the more general, spherically
symmetric, diagonal form of the metric used in problems of gravitational contraction
and expansion [1--4], which we write as
\begin{equation}
ds^2=e^{2\Phi/c^2}c^2dt^2-e^\lambda dr^2-R^2d\Omega^2,
\end{equation}
where, for simplicity, we have introduced the notation $d\Omega^2\equiv d\theta^2+
\sin^2\theta\,d\phi^2$. Here, $\Phi$, $\lambda$, and $R$ are each functions of
$r$ and $t$, and are to be determined by solving Einstein's equations,
\begin{equation}
G_{\alpha\beta}\equiv {\cal R}_{\alpha\beta}-{1\over 2}g_{\alpha\beta}{\cal R}=
-{8\pi G\over c^4}T_{\alpha\beta}\;,
\end{equation}
where ${\cal R}_{\alpha\beta}$ and ${\cal R}$ are the Ricci tensor and scalar,
respectively.

These equations have been solved many times in the literature, so we will
simply borrow the principal results, especially those of Refs. [3,4].
Throughout this paper, an overdot signifies
differentiation with respect to $t$, and a prime indicates differentiation with respect
to $r$.  In addition, we will introduce the so-called Misner--Sharp mass $m(r,t)$, defined as
\begin{equation}
e^{\lambda(r,t)}=g_{rr}=\left[1+U^2-{2Gm(r,t)\over c^2 R}\right]^{-1}\left(R^\prime\right)^2\;,
\end{equation}
where the quantity
\begin{equation}
U\equiv {e^{-\Phi/c^2}\over c}\dot{R}
\end{equation}
gives the relative velocity $U\,d\theta$ (in units of $c$) of adjacent fluid particles on the
same sphere of constant $r$ [3,4,17,18].

We emphasize that we have chosen to work with a system of coordinates moving at each
point with the fluid at that point, a condition first highlighted in Ref. [3].
In this co-moving (or Lagrangian) frame, the four-velocity components are
\begin{equation}
u_t=ce^{-\Phi/c^2}\;,\qquad u_i=0\;\;\;(i=r,\theta,\phi)\;.
\end{equation}
Therefore, the coordinate $t$ must be the time in this co-moving frame, a situation
that contrasts with the more typical approach in which the coordinates are chosen
arbitrarily to simplify the metric before Einstein's equations are invoked 
to determine its coefficients. In such cases, the coordinates are often interpreted
after the solution has been found. However, we are not free to do this here, because the
physical meaning of $t$ has already been employed
to write Eq.~(8). It is therefore straightforward (if somewhat tedious) to
confirm that Einstein's equations result in the following relations for $m(r,t)$:
\begin{equation}
\dot{m}c^2=-4\pi R^2\dot{R}\,p\;,
\end{equation}
and
\begin{equation}
m^\prime c^2=4\pi \rho R^2 R^\prime\;.
\end{equation}
It follows, therefore, that the quantity
\begin{equation}
m(r,t)=\int_0^r {4\pi\over c^2} \rho R^2 R^\prime\,dr
\end{equation}
denotes the mass from the origin (where the observer is located)
out to $r$ at time $t$. Eq.~(9) is the energy equation for the
rate of work due to the pressure.$^*$\footnote{$^*$Incidentally, $m(r,t)$
is also the mass used to define the gravitational horizon
$R_{\rm h}\equiv 2Gm/c^2$ associated with the FRW metric [8],
and it is not difficult to show from this that $R_{\rm h}=c/H$, where
$H\equiv {\dot{a}/ a}$ is the Hubble constant. In other words,
the gravitational horizon defined in terms of the Misner--Sharp mass
is in fact the Hubble radius.} In situations where $m(r_{\rm s},t)$ represents
the mass of a body undergoing gravitational collapse or expansion,
$r_{\rm s}$ would be the radius at its surface, where an important boundary 
condition is $p=0$ [3]. In the cosmological context, Birkhoff's theorem and 
its corollary allow us to consider $m(r,t)$ to be the mass-energy bounded
by a shell of radius $r$ anywhere in the medium, and to view this $m(r,t)$
(and its associated pressure; more on this below) as being solely responsible
for any gravitational influence between the origin and a particle at that radius [5,7,8].

Three more equations are critical to the discussion in this paper. The
first two come from the conservation equation ${T^{\alpha\beta}}_{;\beta}=0$,
which yields the Euler equation
\begin{equation}
{\partial\Phi\over\partial r}=-{p^\prime c^2\over \rho+p}\;
\end{equation}
and the conservation of energy
\begin{equation}
\dot{\rho}=-3\left({\dot{R}\over R}\right)\left(\rho+p\right)\;.
\end{equation}
The dynamical equation may be written as 
\begin{eqnarray}
e^{-\Phi/c^2}{\partial \over\partial t}\left(e^{-\Phi/c^2}\dot{R}\right)
&\hskip-0.3in=\hskip-0.3in&-c^2\left[{1+U^2-2Gm/c^2R\over \rho+p}\right]\times\nonumber\\
&\null&\hskip-0.5in \left({\partial p\over\partial R}\right)_t
-{(Gmc^2+4\pi GR^3p)\over c^2R^2}\;.
\end{eqnarray}
It should be emphasized that these expressions are completely
general for any spherically symmetric distribution of mass-energy
described as a perfect fluid. We have not yet introduced the key
symmetries leading to the metric given in Eq.~(1). In the
following section, we will examine what happens in the cosmological
expansion and stellar collapse scenarios. Specifically, we will see what
is required to reduce the general metric in Eq.~(4) to the more
streamlined FRW formulation of Eq.~(1).

\vspace*{6.0mm} \hrule\vspace{2mm} \noindent {\large
\usefont{T1}{fradmcn}{m}{n}\xbt 3\quad Discussion}\vspace{2.5mm}

\vspace*{5mm} \noindent {3.1\quad Gravitational Collapse}\vspace{3.5mm}

\noindent Let us now begin to introduce some of the principal symmetries.
Suppose the medium is static, so $\dot{R}=0$. Then Eq.~(14)
gives the pressure gradient $\partial p/\partial R$ required to maintain
equilibrium against gravitational collapse. In the special case when
$\rho$ is uniform throughout the sphere, Einstein's equations
show that $\partial/\partial R=e^{-\lambda/2}(\partial/\partial r)_t$,
so we may combine Eqs.~(12) and (14) to arrive at the well-known 
Tolman--Oppenheimer--Volkoff equation [19,20] describing the hydrostatic interior of a star:
\begin{equation}
{\partial\Phi\over\partial R}={Gm(R)+4\pi GR^3p/c^2\over
R^2-2Gm(R)R/c^2}\;.
\end{equation}
In this application of the metric in Eq.~(4), gravity is clearly
present because it counteracts the pressure gradient, and we
see that $\Phi^\prime\not= 0$, which means that $e^{2\Phi/c^2}$
cannot be absorbed into a rescaled time coordinate that would
have allowed us to write $g_{tt}=1$. We would arrive at similar
conclusions should the star be undergoing gravitational collapse, except that
in this case $\dot{R}$ and $U$ are not zero, so the pressure gradient 
would be insufficient to prevent at least partial conversion of gravitational
energy into kinetic energy during the infall.

\vspace*{5mm} \noindent {3.2\quad The Lapse Function in FRW}\vspace{3.5mm}

\noindent Turning now to cosmology, we see that in order to convert the
general metric of Eq.~(4) into the standard FRW form shown
in Eq.~(1), it is necessary to force the condition
\begin{equation}
{\partial\Phi(r,t)\over\partial r}=0\;.
\end{equation}
To do this, however, we must have $p^\prime=0$, which confirms that
the pressure is homogeneous (as well as isotropic), and the dynamical Eq.~(14)
reduces to
\begin{equation}
e^{-\Phi/c^2}{\partial\over\partial t}\left(e^{-\Phi/c^2}\dot{R}\right)=
-{(Gmc^2+4\pi GR^3p)\over c^2R^2}\;,
\end{equation}
or
\begin{equation}
e^{-\Phi/c^2}\dot{R}{\partial\over\partial t}\left(e^{-\Phi/c^2}\dot{R}\right)=
-{Gm\dot{R}\over R^2}-{4\pi G\over c^2}Rp\dot{R}\;.
\end{equation}
From Eq.~(9), the last term on the right-hand side is just $G\dot{m}/R$, and
changing the variable to $u\equiv e^{-\Phi/c^2}\dot{R}$ then gives
\begin{equation}
u{\partial u\over\partial t}={d\over dt}{Gm\over R}\;,
\end{equation}
whose solution is
\begin{equation}
{1\over 2}u^2-{Gm\over R}=K(r)\;,
\end{equation}
where $K(r)$ is an arbitrary function of $r$ only. Anticipating the meaning
of this function in relation to the spatial curvature constant in FRW, we
define $K(r)\equiv -(c^2/2)\tilde{k}(r)r^2$, where $\tilde{k}$ is possibly
a function of $r$, but not of $t$. Thus, the solution to the dynamics
equation in the cosmological context may be written as 
\begin{equation}
{1\over 2}\dot{R}^2\,e^{-2\Phi/c^2}-{Gm\over R}=-{c^2\over 2}\tilde{k}(r)r^2\;.
\end{equation}
Those familiar with FRW dynamics will immediately recognize that this
expression reduces to the Friedmann equation if we impose the constraint
$e^{-2\Phi/c^2}=1$ and the final required symmetry---that $\rho$ should be
uniform throughout the medium.$^\P$\footnote{$^\P$Note also that the inclusion of
Eq.~(19) in Eq.~(6) quite trivially reproduces the FRW form of the
metric coefficient $g_{rr}$.} The $G_{tr}$ component of the Einstein
equations, together with Eq.~(12), would then force $R$ to
have the form $a(t)f(r)$ [4], and the arbitrary function
$f(r)$ could be used to rescale the coordinate $r$ and allow us to recover
Eq.~(1). Therefore, the factor $\tilde{k}$ is indeed the spatial curvature
constant $k$ in the FRW metric, affirming the view that it merely
represents the local energy of the expanding cosmic fluid. Note also
that the factor $r^2$ is common to all the terms in this expression,
so $\Phi$ depends only on the co-moving time $t$, which is consistent with
Eq.~(16).
Equation~(21) couples $\Phi$ to $Gm/R$. This is not
surprising in light of Birkhoff's theorem and its corollary, which indicate that
$Gm/R$ should represent the gravitational potential
on a sphere at $R$ relative to the origin. The lapse function exists specifically
because of the time dilation effects due to the curvature of spacetime.
We are therefore not permitted to {\it arbitrarily} set $e^{2\Phi/c^2}$
equal to 1. We will now formally derive the general expression for
$e^{2\Phi/c^2}$ and show that it is constant only when $\rho+3p=0$.

Because the density $\rho$ and pressure $p$ are uniform,  $g_{rr}$ may be written
as $e^\lambda=a(t)^2$, where the expansion factor $a(t)$ is a function only of
time [4]. Moreover, we showed above that under
these conditions, $\sqrt{g_{\theta\theta}}=R(r,t)=a(t)f(r)$. [As is well known,
the precise form of the function $f(r)$ depends on the value of the spatial
curvature constant $k$ in the FRW metric, which we define below.] Equation~(21)
may therefore be written as
\begin{equation}
\left({\dot{a}\over a}\right)^2={8\pi G\over 3c^2}\rho e^{2\Phi/c^2}-
{kc^2\over a^2}e^{2\Phi/c^2},
\end{equation}
which, as we have already noted, is simply the familiar Friedmann equation,
except for the factor $e^{2\Phi/c^2}$, and we have also defined
$k\equiv \tilde{k}(r)(r/f[r])^2$. From the $rr$ component of the
spherically symmetric perfect-fluid field equations [21] [or
even just simply from Eq.~(17)], one can also easily derive the
corresponding acceleration equation:
\begin{equation}
\left({\ddot{a}\over a}\right)-\left({\dot{a}\over a}\right)
{\dot{\Phi}\over c^2}=-{4\pi G\over 3c^2}e^{2\Phi/c^2}\left(\rho+3p\right)\;.
\end{equation}

We will examine the impact of $\Phi$ on these two equations by first
finding its limiting form when $\rho+3p\rightarrow 0$; i.e., we seek a solution to
Eqs.~(22) and (23) for
\begin{equation}
\rho+3p=\epsilon\rho\;,
\end{equation}
where $\epsilon<<1$.
For this simplified approach, we shall also set $k=0$ (though we relax this
condition for the more general derivation that follows). A straightforward
manipulation of Eq.~(23) then produces the expression
\begin{equation}
{1\over a\dot{a}}{\partial\over\partial t}\left({\dot{a}}^2
e^{-2\Phi/c^2}\right)=-{8\pi G\over 3c^2}\epsilon\rho\;,
\end{equation}
and combining this with Eq.~(22) yields 
\begin{equation}
{\partial\over\partial t}\left\{\ln\left({\dot{a}}^2 e^{-2\Phi/c^2}\right)\right\}
=-\epsilon{\partial\over\partial t}\ln{a}\;.
\end{equation}
The solution to this equation is therefore
\begin{equation}
e^{2\Phi/c^2}=h{\dot{a}}^2a^{\epsilon}\;,
\end{equation}
where $h$ is an integration constant. Clearly, because $a(t)$ is a function of $t$,
$\Phi$ cannot in general be equal to zero. To obtain a time-independent $\Phi$, we must
take the limit $\epsilon\rightarrow 0$, in which case $\Phi$ would be constant
as long as $\dot{a}$ is also constant.

The more general form of Eq.~(26), without the use of Eq.~(24)
(and without setting $k=0$), is
\begin{eqnarray}
{\partial\over\partial t}\left\{\ln\left({\dot{a}}^2 e^{-2\Phi/c^2}\right)\right\}
&\hskip-0.3in=\hskip-0.3in&-{kc^2\over a\dot{a}}\left(1+{3p\over \rho}\right)e^{2\Phi/c^2}-\nonumber\\
&\null&\hskip-0.5in{\dot{a}\over a}\left(1+{3p\over\rho}\right)\;.
\end{eqnarray}
The solution to this equation may be written as 
\begin{equation}
e^{2\Phi(t)/c^2}=h{\dot{a}}^2e^{{\cal I}(t)}\;,
\end{equation}
with
\begin{equation}
{\cal I}(t)\equiv \int^t_0 dt^\prime\;{8\pi G\over 3c^2H}e^{\Phi/c^2}(\rho+3p)\;,
\end{equation}
where $H\equiv e^{-\Phi/c^2}(\dot{a}/a)$ is the Hubble constant, and the
integrand is a function of $t^\prime$ only. This expression is
more complicated than Eq.~(27), but the result is the same.
To achieve a constant $\Phi$, we must have ${\cal I}\rightarrow 0$,
which is guaranteed only when $\rho+3p\rightarrow 0$. Then $\Phi$ is
constant as long as $\dot{a}$ is constant [which is ensured by Eq.~(23)],
and the lapse function may be set equal to 1 with an appropriate choice of
the initial condition $h$.

\vspace*{5mm} \noindent {3.3\quad Uniqueness of the Co-moving, Free-falling Frame}\vspace{3.5mm}

\noindent Suppose we were to choose a cosmic equation of state such that $\rho+3p\not=0$.
In that case, we know that $\Phi(t)$ cannot be constant, though it is a
function only of $t$, not of the spatial coordinates. The reason for this is clear. In other
spherically symmetric spacetimes, such as the Schwarzschild spacetime, where the curvature
depends on the position, the time dilation is itself a function of $r$ (though for
that particular spacetime the curvature is static, so $\Phi$ is independent
of time). This results in a spatially dependent lapse function to which we
are accustomed. In the FRW metric, however, the Universe is homogeneous
and isotropic throughout each time slice, so the lapse function $e^{2\Phi/c^2}$
must be independent of $(r,\theta,\phi)$; it can change only from slice to
slice if the spacetime curvature is evolving with time.

The fact that $\Phi$ is a function only of $t$ could be viewed as an
inconsequential ``gauge'' freedom. In other words, the interpretation of
$e^{2\Phi/c^2}$ as a true lapse function representing a spatially uniform
time dilation in this metric would not be recognized as such. However, now that we have
formally derived the metric coefficients in FRW from the general form
of the spherically symmetric metric, we can demonstrate that the supposed
change in gauge, $g_{tt}\rightarrow 1$, is actually a transformation of
the coordinates into the free-falling frame. It is only in this frame
that $dt$ can reduce to the usual (local) proper time $d\tau\equiv 
ds/c$, so $g_{tt}=1$ (corresponding to an acceleration-free environment).

The required coordinate transformation to eliminate the lapse function in
Eq.~(4) is
\begin{equation}
d\tilde{t}\equiv e^{\Phi(t)/c^2}\,dt\;,
\end{equation}
so 
\begin{equation}
\tilde{t}=\int_0^\tau e^{\Phi(t^\prime)/c^2}\,dt^\prime\;.
\end{equation}
The coordinate $\tilde{t}$ therefore subsumes the accumulated
(spatially uniform) dilation of $t$ when $\Phi\not= 0$.
Correspondingly, if we were to also define
$\dot{\tilde{a}}\equiv da(\tilde{t})/d\tilde{t}$, then
\begin{equation}
\dot{\tilde{a}} =  e^{-\Phi(t)/c^2}\,\dot{a},
\end{equation}
and rewriting Eqs.~(22) and (23) in terms of $\tilde{a}$,
$\dot{\tilde{a}}$, and $\ddot{\tilde{a}}$ would then recover the
familiar Friedmann and acceleration equations, though with the
derivatives now written in terms of $\tilde{t}$ rather than $t$.
As expected, this transformation has placed us in the free-falling
frame, corresponding to the FRW metric in Eq.~(1), with the time
coordinate now given as $\tilde{t}$.

However, we already selected our set of coordinates to
be those in the co-moving (i.e., Lagrangian) frame from the beginning.
This was necessary to derive our equations, starting with
the four-velocity in Eq.~(8), which allowed us to move with the
fluid at each spacetime point. The transformation in Eq.~(31)
to put us in the free-falling frame, where $d\tilde{t}\rightarrow d\tau$,
therefore does not represent a true gauge freedom at all, because
in the context of FRW, the free-falling and co-moving frames
are one and the same. If we want to recover the FRW metric
in Eq.~(1), the choice of gauge is not free because
the uniqueness of the co-moving and free-falling frames
forces $t$ and $\tilde{t}$ ($=\tau$) to be the same coordinate.
$\Phi(t)$ must always be identically zero, so, from Eq.~(30), we
must have $\rho+3p\rightarrow 0$.

As a concrete example, consider what happens to an Einstein--de Sitter
spacetime under such a transformation. Written in terms of $\tilde{t}$,
the expansion factor in a Universe containing only matter (with
corresponding zero pressure) has the well-known solution $a(\tilde{t})=
{\tilde{t}}^{2/3}$. However, because $\rho+3p\not= 0$ in this case,
$a(t)\not= t^{2/3}$. Conceptually, a gauge transformation is supposed
to leave the equations of motion unchanged, yet here, the geodesics written
in $t$ are different from those written in $\tilde{t}$, even though we
are using the same $\rho$ and $p$.

The FRW metric is special among spherically symmetric spacetimes because of
its elegance and simplicity. However, its attractiveness and practicality come at a
cost---they are valid only for a perfect fluid whose equation of state is uniquely
given by the expression $p=-\rho/3$ and whose expansion rate is therefore
constant, with $\ddot{a}=0$.

\vfill\newpage
\vspace*{6.0mm} \hrule\vspace{2mm} \noindent {\large
\usefont{T1}{fradmcn}{m}{n}\xbt 4\quad Conclusion}\vspace{2.5mm}

\noindent Current cosmological observations are precise enough to test whether
this conclusion is confirmed in reality. Despite the perception that this
result may be in conflict with these measurements, quite the opposite is
true. The zero active mass condition gives rise to what we have been
calling the $R_{\rm h}=ct$ Universe in the literature [8,22,23].
As the quality of the observations
continues to improve, we increasingly see that optimization of the
parameters in $\Lambda$CDM/$w$CDM brings the overall expansion history
in this model ever closer to that expected in a Universe with $p=-\rho/3$.
The evidence comes from cosmic chronometers [24,25], gamma-ray bursts 
[26,29], high-redshift quasars [30], Type Ia SNe [31], and, most recently, 
an application of the Alcock--Paczy\'nski test using model-independent baryon 
acoustic oscillation (BAO) data [32-34], among others.
The BAO measurements are particularly noteworthy because, with their
$\sim4\%$ accuracy, they now rule out the standard model when the zero active
mass condition is ignored at better than the $99.34\%$ C.L. Instead, they strongly
favor the $R_{\rm h}=ct$ model, with its equation of state $p=-\rho/3$.
The conclusion from comparative studies such as these is that, although
$\Lambda$CDM is a parameter-rich cosmology, ultimately, when its
parameters are optimized to fit the data, its predictions fall in line with
the expansion history we would have expected all along in an FRW
spacetime for a fluid with zero active mass.

\vskip 0.2in
\noindent {\baselineskip
10.5pt\renewcommand{\baselinestretch}{1.05}\footnotesize {\bf
Acknowledgements}~~~Part of this research was carried out
at the International Space Science Institute in Bern, Switzerland, and under
the sponsorship of the Simpson visiting chair at Amherst College..}\vspace{7mm}

\hrule\vspace{2mm}

\noindent {\large
\usefont{T1}{fradmcn}{m}{n}\bf References}\vspace{3.1mm}
\parskip=0mm \baselineskip 15pt\renewcommand{\baselinestretch}{1.25} \footnotesize
\parindent=9mm
\setlength{\baselineskip}{12.3pt}   

\newcommand{\aj}{AJ}
\newcommand{\apj}{ApJ}
\newcommand{\apjs}{ApJS}
\newcommand{\apjl}{ApJ}
\newcommand{\mnras}{MNRAS}
\newcommand{\pasj}{PASJ}
\newcommand{\aap}{A\&A}
\newcommand{\nat}{Nature}
\newcommand{\na}{New Astr.}
\newcommand{\araa}{ARA\&A}
\newcommand{\apss}{Ap\&SS}
\newcommand{\prd}{Phys. Rev. D}
\newcommand{\jcap}{JCAP}
\noindent {\bf [1] } J. R. Oppenheimer and H. Snyder, On continued gravitational
contraction, {\it Phys. Rev.} 56(5), 455 (1939)
\\[1mm] {\bf [2] } G. C. McVittie, Gravitational collapse to a small volume, {\it Astrophys. J.} 140, 401 (1964)
\\[1mm] {\bf [3] } C. W. Misner and D. H. Sharp, Relativistic equations for adiabatic,
spherically symmetric gravitational collapse, {\it Phys. Rev.} 136(2B), B571 (1964)
\\[1mm] {\bf [4] } I. H. Thompson \& G. F. Whitrow, Time-dependent internal
solutions for spherically symmetrical bodies in general relativity (I): Adiabatic collapse,
{\it Mon. Not. R. Astron. Soc.} 136(2), 207 (1967)
\\[1mm] {\bf [5] } G. Birkhoff, Relativity and Modern Physics, Harvard University Press 1923
\\[1mm] {\bf [6] } H. P. Robertson, On the foundations of relativistic cosmology,
{\it Proc. Natl. Acad. Sci. USA} 15(11), 822 (1929)
\\[1mm] {\bf [7] } S. Weinberg, Gravitation and Cosmology: Principles and Applications
of the General Theory of Relativity, Wiley 1972
\\[1mm] {\bf [8] } F. Melia, The cosmic horizon, {\it Mon. Not. R. Astron. Soc.} 382(4), 1917 (2007)
\\[1mm] {\bf [9] } D. H. Weinberg, M. J. Mortonson, D. J. Eisenstein, C. Hirata, A. G. Riess,
and E. Rozo, Observational probes of cosmic acceleration, {\it Phys. Rep.} 530(2), 87 (2013)
\\[1mm] {\bf [10] } S. Perlmutter, G. Aldering, G. Goldhaber, R. A. Knop, P.  Nugent, P. G. Castro, 
S. Deustua, S. Fabbro, A. Goobar, D. E. Groom, I. M. Hook, A. G. Kim, M. Y. Kim, J. C. Lee, N. J. Nunes, 
R. Pain, C. R. Pennypacker, R. Quimby, C. Lidman, R. S. Ellis, M. Irwin, R. G. McMahon, P. Ruiz-Lapuente, 
N. Walton, B. Schaefer, B. J. Boyle, A. V. Filippenko, T. Matheson, A. S. Fruchter, N. Panagia, 
H. J. M. Newberg, W. J. Couch, and T. S. C. Project, Measurements of W and L from 42 high-redshift 
supernovae, {\it Astrophys. J.} 517(2), 565 (1999)
\\[1mm] {\bf [11] } A. G. Riess, A. V. Filippenko, P. Challis, A. Clocchiatti, A.
Diercks, P. M. Garnavich, R. L. Gilliland, C. J. Hogan, S.
Jha, R. P. Kirshner, B. Leibundgut, M.M. Phillips, D. Reiss, B. P. Schmidt, R. A. Schommer, R. C. Smith, J. Spyromilio,
C. Stubbs, N. B. Suntzeff, and J. Tonry, Observational evidence from supernovae for an accelerating universe and a
cosmological constant, {\it Astron. J.} 116(3), 1009 (1998)
\\[1mm] {\bf [12] } M. Kowalski, D. Rubin, G. Aldering, R. J. Agostinho, A.
Amadon, et al., Improved cosmological constraints from new, old, and combined supernova data sets,
{\it Astrophys. J.} 686(2), 749 (2008)
\\[1mm] {\bf [13] } N. Suzuki, D. Rubin, C. Lidman, G. Aldering, R. Amanullah,
et al., The Hubble space telescope cluster supernova survey
(v): Improving the dark-energy constraints above z > 1
and building an early-type-hosted supernova sample, {\it Astrophys. J.} 746(1), 85 (2012)
\\[1mm] {\bf [14] } C. L. Bennett, R. S. Hill, G. Hinshaw, M. R. Nolta, N. Odegard,
L. Page, D. N. Spergel, J. L.Weiland, E. L. Wright, M. Halpern, N. Jarosik, A. Kogut, 
M. Limon, S. S. Meyer, G. S. Tucker, and E. Wollack, First–Year Wilkinson Microwave
Anisotropy Probe (WMAP) Observations: Foreground emission, {\it Astrophys. J. Suppl.} 148(1), 97 (2003)
\\[1mm] {\bf [15] } D. N. Spergel, L. Verde, H. V. Peiris, E. Komatsu, M. R.
Nolta, C. L. Bennett, M. Halpern, G. Hinshaw, N. Jarosik,
A. Kogut, M. Limon, S. S. Meyer, L. Page, G. S. Tucker,
J. L. Weiland, E. Wollack, and E. L. Wright, First–Year
Wilkinson Microwave Anisotropy Probe (WMAP) Observations:
Determination of cosmological parameters, {\it Astrophys. J. Suppl.} 148(1), 175 (2003)
\\[1mm] {\bf [16] } P. A. R. Ade, et al. (Planck Collaboration), Planck 2013 results
(XXIII): Isotropy and statistics of the CMB, {\it Astron. Astrophys.} 571, A23 (2014)
\\[1mm] {\bf [17] } W. C. Hernandez and C. W. Misner, Observer time as a coordinate
in relativistic spherical hydrodynamics, {\it Astrophys. J.} 143, 452 (1966)
\\[1mm] {\bf [18] } M. M. May and R. H. White, Hydrodynamic calculations of
general-relativistic collapse, {\it Phys. Rev.} 141(4), 1232 (1966)
\\[1mm] {\bf [19] } R. C. Tolman, Static solutions of Einstein’s field equations
for spheres of fluid, {\it Phys. Rev.} 55(4), 364 (1939)
\\[1mm] {\bf [20] } J. R. Oppenheimer and G. M. Volkoff, On massive neutron
cores, {\it Phys. Rev.} 55(4), 374 (1949)
\\[1mm] {\bf [21] } H. Stephani, D. Kramer,  M. MacCallum and C. Hoenselaers,  Exact
Solutions to Einstein's Field Equations, Cambridge University Press 2009
\\[1mm] {\bf [22] } F. Melia and M. Abdelqader, The cosmological spacetime,
{\it Int. J. Mod. Phys. D} 18(12), 1889 (2009)
\\[1mm] {\bf [23] } F. Melia and A. Shevchuk, The $R_{\rm h} = ct$ universe,
{\it Mon. Not. R. Astron. Soc.} 419(3), 2579 (2011)
\\[1mm] {\bf [24] } R. Jimenez and A. Loeb, Constraining cosmological parameters
based on relative galaxy, {\it Astrophys. J.} 573(1), 37 (2002)
\\[1mm] {\bf [25] } F. Melia and R. S. Maier, Cosmic chronometers in the
$R_{\rm h} = ct$ universe, {\it Mon. Not. R. Astron. Soc.} 432(4), 2669 (2013)
\\[1mm] {\bf [26] } B. E. Schaefer, Gamma-ray burst Hubble diagram to z = 4.5,
{\it Astrophys. J.} 583(2), L67 (2003) 
\\[1mm] {\bf [27] } G. Ghirlanda, G. Ghisellini and D. Lazzati, The collimation-corrected
gamma-ray burst energies correlate with the peak energy of their $\nu F_\nu$ spectrum,
{\it Astrophys. J.} 616(1), 331 (2004)
\\[1mm] {\bf [28] } E. Liang and B. Zhang, Model-independent multivariable
gamma-ray burst luminosity indicator and its possible cosmological
implications, {\it Astrophys. J.} 633(2), 611 (2005)
\\[1mm] {\bf [29] } J.-J. Wei, X.-F. Wu and F. Melia, The gamma-ray burst
Hubble diagram and its implications for cosmology, {\it Astrophys. J.} 772(1), 43 (2013)
\\[1mm] {\bf [30] } F. Melia, High-z quasars in the $R_{\rm h} = ct$ universe,
{\it Astrophys. J.} 764(1), 72 (2013)
\\[1mm] {\bf [31] } J.-J. Wei, X.-F. Wu, F. Melia and R. S. Maier, A comparative
analysis of the supernova legacy survey sample with ΛCDM and the $R_{\rm h} = ct$ universe,
{\it Astron. J.} 149(3), 102 (2015)
\\[1mm] {\bf [32] } A. Font-Ribera, D. Kirkby, N. Busca, J. Miralda-Escud´e,
N. P. Ross, et al., Quasar-Lyman α forest cross-correlation
from BOSS DR11: Baryon acoustic oscillations, {\it J. Cosmol. Astropart. Phys.} 05, 027 (2014)
\\[1mm] {\bf [33] } T. Delubac,J. E. Bautista, N. G. Busca, J. Rich, D. Kirkby,
et al., Baryon acoustic oscillations in the Lya forest of BOSS DR11 quasars,
{\it Astron. Astrophys.} 574, A59 (2015), arXiv:1404.1801 
\\[1mm] {\bf [34] } F. Melia and M. Lop\'ez Corredoira, Alcock–Paczynski test with
model-independent BAO Data, {\it Mon. Not. R. Astron. Soc.} (submitted), arXiv:1503.05052, 2015

\end{multicols}
\end{document}